\documentclass[
twocolumn,
aps,prl,
nofootinbib,
superscriptaddress,
showpacs,
tightenlines,
]{revtex4}
\usepackage{amsmath}
\usepackage{amssymb}
\usepackage{bm}
\usepackage{color,graphicx}
\usepackage{subfigure}

\newcommand{\ba}{\begin{eqnarray}}
\newcommand{\ea}{\end{eqnarray}}
\newcommand{\eT}{\epsilon_T}

\begin{document}

\title{Beyond-Standard-Model Tensor Interaction and Hadron Phenomenology}

\author{Aurore Courtoy} 
\email{aurore.courtoy@ulg.ac.be}
\affiliation{IFPA, AGO Department, Universit\'e de Li\`ege, B\^at. B5, Sart Tilman B-4000 Li\`ege, Belgium
\\ and Divisi\'on  de Ciencias e Ingenier\'ias, Universidad de Guanajuato, C.P. 37150, Le\'on, M\'exico }

\author{Stefan Baessler}
\email{baessler@virginia.edu}
\affiliation{University of Virginia - Physics Department,
382 McCormick Rd., Charlottesville, Virginia 22904 - USA\\
and Oak Ridge National Laboratory, Oak Ridge, TN 37831, USA. } 

\author{Mart\'in Gonz\'alez--Alonso}
\email{mgonzalez@ipnl.in2p3.fr}
\affiliation{IPNL, Universit\'e de Lyon, Universit\'e Lyon 1, CNRS/IN2P3, 
4 rue E. Fermi 69622 Villeurbanne Cedex, France.} 

\author{Simonetta Liuti }
\email{sl4y@virginia.edu}
\affiliation{University of Virginia - Physics Department,
382 McCormick Rd., Charlottesville, Virginia 22904 - USA \\ and Laboratori Nazionali di Frascati, INFN, Frascati, Italy.}


\begin{abstract}
We evaluate the impact of recent developments in hadron phenomenology on extracting possible fundamental tensor interactions beyond the standard model. 
We show that a novel class of  observables, including the chiral-odd generalized parton distributions, and the transversity parton distribution function 
can contribute to the  constraints on this quantity. Experimental extractions of the tensor hadronic matrix elements, if sufficiently precise, will provide a so far absent testing ground for lattice QCD calculations. 
 \end{abstract}

\maketitle

\baselineskip 3.0ex

High precision measurements of beta decay observables play an important role in beyond the standard model (BSM) physics searches,  as they allow us to probe couplings  other than of the $V-A$ type, which could appear at the low energy scale. 
Experiments using cold and ultra-cold neutrons \cite{Abele:2008zz,Nico:2009zua,Young:2014mxa,Baessler:2014gha}, nuclei \cite{Gonzalez-Alonso:2013uqa,Severijns:2014iha,Hardy:2013lga,Behr:2014hha},  and meson rare decays  \cite{Pocanic:2014jka}, are being performed, or have been planned,  that can reach the per-mil level or even higher precision.
Effective field theory (EFT)  allows one to connect these measurements  and BSM effects generated at TeV scales. In this approach that  complements collider searches,
the new  interactions are introduced in an effective Lagrangian describing semi-leptonic transitions at the GeV scale including  four-fermion terms,
or operators up to dimension six  for the scalar, tensor, pseudo-scalar, and V+A interactions (for  a review of the various EFT approaches see Ref.\cite{Cirigliano:2013xha}).
Because the strength of the new interactions is defined with respect to the strength of the known SM interaction, 
the coefficients of the various terms, $\epsilon_i$,  ($i=S,T,P,L,R$) 
depend on the ratio $m_W^2/\Lambda_i^2$, where $\Lambda_i$ is the new physics scale relevant for these non-standard interactions, and $m_W^2$ enters through $G_F=g^2/(4\sqrt{2}m_W^2)$. 
Therefore, the precision with which  $\epsilon_i \propto m_W^2 / \Lambda_i^2$, is known determines a lower limit for $\Lambda_i$.
%
%
The scalar (S) and tensor (T) operators, in particular,  contribute linearly  to the beta decay parameters through their interference with the SM amplitude, and they are therefore more easily detectable. 
The matrix elements/transition amplitudes  between neutron and proton states of all quark bilinear Lorentz structures in the effective Lagrangian which are relevant for  beta decay observables, 
involve products of the BSM  couplings, $\epsilon_i$,  and the corresponding hadronic charges, $g_i$, {\it i.e.} considering  only terms with left-handed neutrinos, 
$C_S =   G_F V_{ud} \sqrt{2} \epsilon_{S}  g_{S}$, and $C_T =   4 G_F V_{ud} \sqrt{2} \epsilon_{T}  g_{T}$.
$g_{S(T)}$ can be parameterized in terms of nucleon form factors which cannot be measured directly, being chiral odd.

%
Various approaches have been  developed so far to calculate these quantities
including lattice QCD \cite{Gockeler:2006zu,Green:2012ej,Bhattacharya:2013ehc,Bali:2014nma,Gonzalez-Alonso:2013ura}, and most recently Dyson-Schwinger Equations \cite{Yamanaka:2013zoa,Pitschmann:2014jxa}. 
Lattice QCD provides the most reliably calculated values for the isovector scalar and tensor charges with precision levels of $\Delta g_S/g_S \approx 15\%$, and $\Delta g_T/g_T \lesssim 4 \%$, respectively. 
Following the analysis in Ref.\cite{Bhattacharya:2011qm}, these values are well below the minimum accuracy that is required not to deteriorate the per mil level  constraints from decay experiments.

In this Letter we call attention to the fact that the  
nucleon form factors  which are relevant for BSM physics searches using neutron beta decay at the scale $t=(M_n-M_p)^2 \approx 0$, can now also be measured accurately in deep inelastic processes that occur at  the multi-GeV scale.  
This novel development emerges from 
recent experimental and theoretical advances in the study of the 3D structure of the nucleon. We focus on $g_T$ that appears at leading order in the hadroproduction cross section, and we evaluate both the uncertainty from the experimental extraction of this quantity, and its impact on the determination of the elementary tensor coupling, $\epsilon_T$.
Current and future planned experiments on dihadron semi-inclusive and deeply virtual exclusive pseudoscalar meson ($\pi^o$ and $\eta$) electroproduction at Jefferson Lab \cite{ourprop_solid,Bedlinskiy:2012be} and COMPASS \cite{Adolph:2012nw,Adolph:2014fjw} allow us to measure  $g_T$ 
with an improved accuracy.    
The main outcome of  the analysis presented here is that the new, more precise measurements of the tensor charge provide for the first time a constraint from experiment on the hadronic matrix element  in BSM searches. 


The tensor form factor is derived from an integral relation involving the transversity  (generalized) parton distribution function, or the probablity to find a quark with a  net transverse polarization in a transversely polarized proton, 
\begin{eqnarray}
\label{formf:eq}
 g_T^q (t,Q^2) \!   =  \! \int_{0}^1 dx    \, \left[ H_T^q(x,\xi,t;Q^2) - H_T^{\bar{q}}(x,\xi,t;Q^2) \right]  &&  \\
\label{charge:eq}
g_T^q (0, Q^2)   =  \int_0^1 dx \, [ h_1^q(x, Q^2) - h_1^{\bar{q}}(x, Q^2)], &&
\end{eqnarray}
where, $h_1^{q(\bar{q})}(x, Q^2)$  \cite{Ralston:1979ys,Jaffe:1992ra}, and $H_T^{q(\bar{q})}(x,\xi,t;Q^2)$ \cite{Diehl:2001pm}, are the quark (anti-quark) transversity Parton Distribution Function (PDF) and Generalized Parton Distribution (GPD), respectively;
$Q^2$ is the virtual photon's four momentum squared in the deep inelastic processes defining each object, while $t = (p- p')^2$ is the four-momentum transfer squared between the initial $(p)$ and final  $(p')$ proton, $t=0$ for a PDF which corresponds to the imaginary part of the forward amplitude;   
$x$ and $\xi$ are parton longitudinal momentum fractions which are connected to $x_{Bj} = Q^2/2M\nu$, $\nu$ being the energy transfer.
%

The occurrence of these types of integral relations in the chiral odd sector parallels in some respect the Bjorken sum rule \cite{Bjorken:1966jh} connecting the nucleon's helicity structure functions and the axial charge. For the tensor form factor and charge, however, given the non renormalizability of the tensor interactions,  current algebra cannot be applied. Notice that the QCD Lagrangian does not allow for a proper conserved current associated to the tensor ``charge" which 
is in itself somewhat a misnomer. In fact, the tensor charge evolves with
the hard scale $Q^2$, in perturbative QCD \cite{Barone:2001sp,Wakamatsu:2008ki}. 

The transversity distributions in Eqs.(\ref{formf:eq},\ref{charge:eq}) parametrize the tensor interaction component in the quark-quark correlation function which reads,
\begin{eqnarray}
\label{eq:transversitySR}
  \int\frac{dz^-}{4\pi} e^{i x \overline{P}^+z^-}\left \langle  p' \,  S'_{\perp}\right| {\overline q}(0)\,  {\mathcal O}_T^\pm  q(z)\left | p  S_{\perp}\right \rangle |_{z^+ = {\bf z}_T=0 }, 
\end{eqnarray}
where $ | p  \, S_{\perp}  \rangle$ represents  the proton's ``transversity state", or a state with transverse polarization obtained from a superposition of states in the  helicity basis; 
the quarks fields ($q=u,d$)  tensor structure,
${\cal O}_T^\pm   =  -i  (\sigma^{+1} \pm i \sigma^{+2})$, is chiral odd or,
it connects quarks with opposite helicities.
By working out the detailed helicity structure of the correlation function, one finds that the relevant combination defining transversity is 
the net transverse quark polarization in a transversely polarized proton.

The isovector components of the tensor hadronic matrix element which are relevant for neutron beta decay correspond to the same tensor structure in Eq.(\ref{eq:transversitySR}),
 taking the quark fields operator to be local, namely, $\bar{q}(0) {\cal O}_T^\pm q(0)$, 
\begin{eqnarray}
\left \langle p_p', S_p' \right | \bar{u} \sigma_{\mu\nu} u - \bar{d} \sigma_{\mu\nu} d  \left | p_p, S_p \right\rangle = 
 g_T(t,Q^2) \, {\overline U}_{p'} \sigma_{\mu\nu}U_p,
\end{eqnarray}
where $g_{T}^q$ represents the tensor form factor for the flavor $q$ in the proton, $g_{T}^u \equiv g_{T}^{u/p}$, and $g_{T}^d \equiv g_{T}^{d/p}$. 
From isospin symmetry one can write,
\begin{eqnarray}
\left \langle p_p, S_p \right | \bar{u} \sigma_{\mu\nu} d  \left | p_n, S_n\right\rangle  &=& g_T(t, Q^2)\, {\overline U}_{p_p} \sigma_{\mu\nu} U_{p_n}, 
\end{eqnarray}
where $p_n \rightarrow p_p$, and  $p_p \rightarrow p_p'$.
%

%

Transversity cannot be measured in an ordinary deep inelastic scattering process  because it is a chiral-odd quantity, but it has been measured with large errors 
in one-pion jet  semi-inclusive deep inelastic scattering (SIDIS) with transversely polarized targets (see review in \cite{Burkardt:2008jw}).
Recent progress in both dihadron SIDIS  and exclusive deeply virtual meson electroproduction (DVMP) experiments  have, however, relaunched the possibility of obtaining a precise experimental determination of $g_T^q$. The main reason why these processes can provide a cleaner measurement  is that they are not sensitive to intrinsic transverse momentum dependent distributions and fragmentation functions, and they therefore connect more directly to the tensor charge while obeying simpler factorization theorems in QCD. 

\begin{center}
\begin{figure}
\includegraphics[width=7.cm]{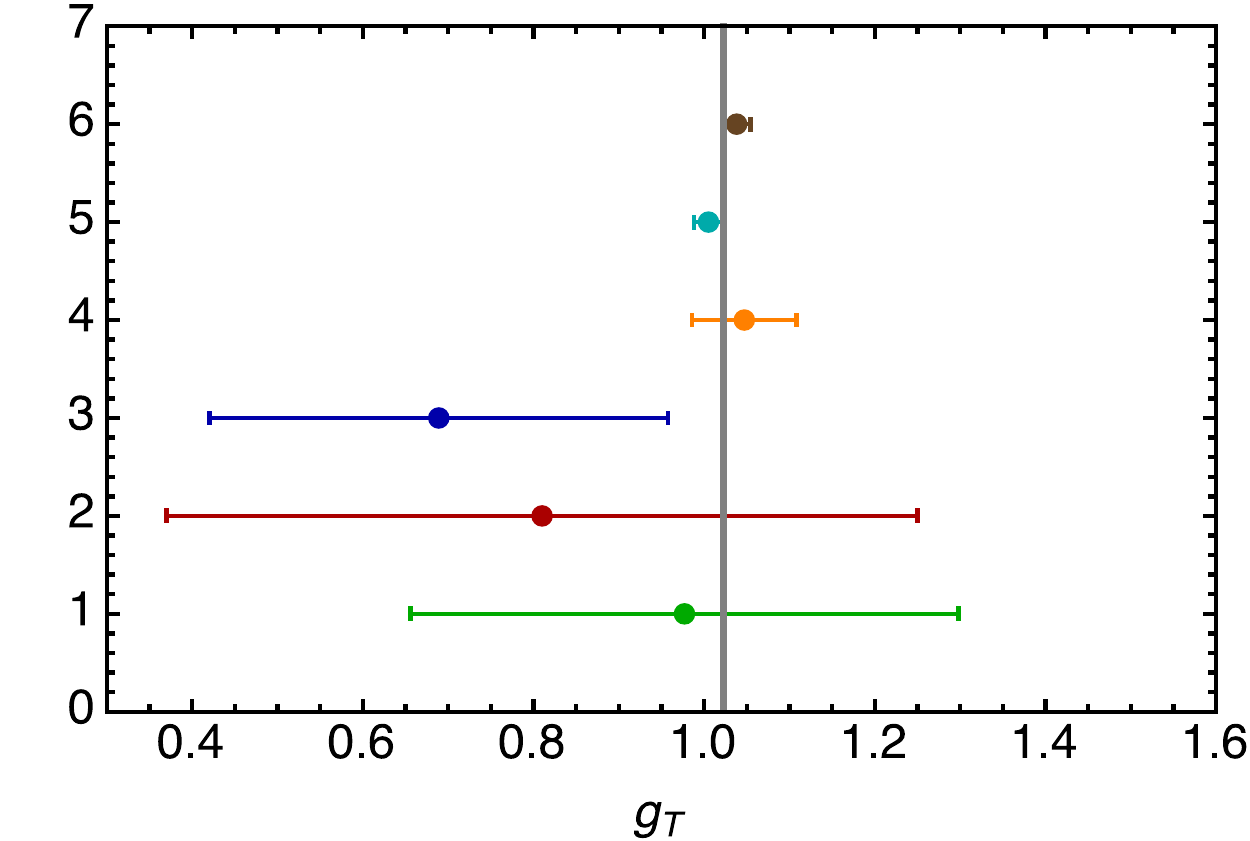}
\caption{(Color online) Values of the tensor charge, $g_T(0,4$GeV$^2)$  with its uncertainty as obtained in: (1) DVMP, Ref.~\cite{Goldstein:2014aja};  (2) {\it flexible form} DiFF, Ref.~\cite{Radici:2015mwa}; (3) Single pion jet SIDIS, Ref.~\cite{Anselmino:2008jk}; Lattice QCD: (4) RQCD \cite{Bali:2014nma}, (5) LHPC~\cite{Green:2012ej},  (6) PNDME ~\cite{Bhattacharya:2013ehc}.
 }
\label{fig:gt}
\end{figure}
\end{center}

Dihadron SIDIS off transversly polarized targets,
\begin{equation}
 l + N \to l' + H_1 + H_2 + X\quad, \nonumber
 \end{equation}
where $l$ denotes the (unpolarized) lepton beam, $N$ the nucleon target, $H_1$ and $H_2$ the produced hadrons,  allows one to access the $h_1$, through  the modulation from the azimuthal angle $\phi_S$ of the target polarization  component $S_T$, transverse to both the virtual-photon and target momenta,
and the azimuthal angle of the transverse average momentum of the pion pair $\phi_R$ {\it w.r.t.} the virtual photon direction.
In this process, 
the observable can be written as the product of $h_1^q$, and a chiral odd fragmentation function called $H_1^{\sphericalangle\, q}$ 
~\cite{Bianconi:1999cd,Radici:2001na},
\begin{eqnarray}
 F_{UT}^{\sin (\phi_R +\phi_S)} & = & \, x\,
\sum_q e_q^2\,  h_1^q(x; Q^2)\,\frac{|\bm R| \sin \theta}{M_h} \, H_1^{\sphericalangle\, q}.
\end{eqnarray}
%
%
Data for the single-spin asymmetry related to the modulation of interest here are available from HERMES~\cite{Airapetian:2008sk} and COMPASS~\cite{Adolph:2012nw,Adolph:2014fjw} on both proton and deuteron target allowing for a $u$ and $d$ quarks flavor separation, whereas the chiral-odd DiFF have been extracted from  the angular distribution of two pion pairs produced in $e^+ e^-$ annihilations at Belle~\cite{Courtoy:2012ry}.
Using these data sets, in Ref.~\cite{Bacchetta:2012ty,Radici:2015mwa},  the transversity PDF has been determined for different functional forms, and using  the {\it replica} method for the error analysis.
As for future extractions, the dihadron SIDIS  will be studied in CLAS12 at JLab on a proton target and in SoLID on a neutron target \cite{ourprop_solid} that will give 
both an improvement  of $\sim 10\%$ in the ratio $\Delta g_T/g_T$ thanks to a wider kinematical coverage and better measurement of the $d$ quarks contribution. 
The results from this extraction are shown in Figure \ref{fig:gt}.
%

Deeply virtual exclusive pseudoscalar meson production (DVMP),
\[ l + N \rightarrow l' + \pi^o (\eta) +  N', \] 
was proposed as a way to access transversity GPDs assuming a (twist three) chiral odd coupling ($\propto \gamma_5$) for  the $\pi^o (\eta)$ prompt production mechanism
 \cite{Ahmad:2008hp,Goldstein:2012az,Goldstein:2013gra,Goldstein:2014aja,Goloskokov:2011rd,Goloskokov:2013mba}.  
%
Three additional transverse spin configurations are allowed in the proton besides transversity which can be described in terms of combinations of GPDs called
$E_T, \widetilde{H}_T, \widetilde{E}_T$ \cite{Diehl:2001pm}. The GPDs enter the observables at the amplitude level, convoluted with  complex coefficients at the leading order, thus forming the generalized form form factors (GFFs).  The various cross section terms and asymmetries are bilinear functions of the GFFs. 
%
A careful analysis of the helicity amplitudes contributing to DVMP has to be performed in order to disentangle the various chiral odd GFFs
from experiment \cite{Kim:2014vea}. 

The ideal set of data to maximally constrain the tensor charge in the chiral odd sector are
from the transverse target spin asymmetry modulation \cite{Goldstein:2014aja},
\begin{eqnarray}
F_{UT}^{\sin(\phi-\phi_S)} & = & 
  \Im m \left[{\cal H}_T^* (2  \widetilde{\cal H}_T + {\cal E}_T) \right] 
 \label{A_UTsinphim} 
\end{eqnarray}
where  $\phi$, is the angle between the leptonic and hadronic planes, and $\phi_s$, the angle between the lepton's plane and the outgoing hadron's transverse spin. 
%
In Ref.\cite{Goldstein:2014aja} the tensor charge was, however, extracted by fitting the unpolarized $\pi^o$ production cross section \cite{Bedlinskiy:2012be}, using a parametrization constrained from data in the chiral even sector to guide the functional shape of the in principle unknown chiral odd GPDs.
Notice that the tensor charge was obtained with a relatively small error because of the presence of these constraints. 
The results from this extraction are also shown  in Fig. \ref{fig:gt}.


Finally, in Fig. \ref{fig:gt} we quote also the value obtained in single pion SIDIS \cite{Anselmino:2008jk}, although this is known to contain some unaccounted for corrections from TMD evolution 
\cite{Collins:2014jpa,Kang:2014zza}. 
%
\begin{center}
\begin{figure}
\includegraphics[width=7.cm]{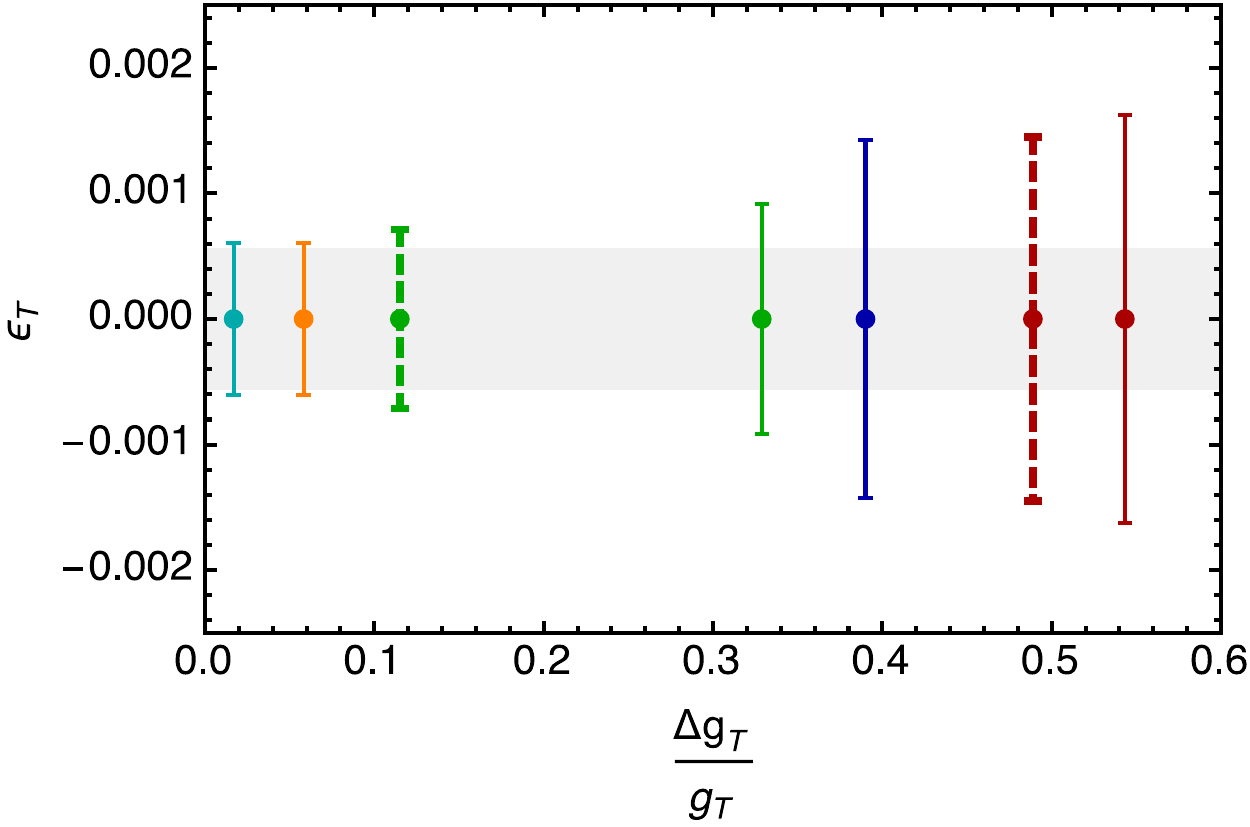}
\caption{(Color online) Bounds on $\eT$  obtained from precision measurements of beta decay using all current extractions and lattice QCD evaluations of the tensor charge $g_T$, plotted vs. the relative error, $\Delta g_T/g_T$ described in the text: (turqoise)  Lattice QCD \cite{Green:2012ej,Bhattacharya:2013ehc}; (yellow) Lattice QCD \cite{Bali:2014nma};
(green) Deeply virtual $\pi^o$ and $\eta$ production \cite{Goldstein:2014aja}; (blue) single pion SIDIS \cite{Anselmino:2008jk};  (red) dihadron SIDIS \cite{Radici:2015mwa}. The dashed lines are future projections.
All results were obtained using in the definition of $\Delta g_T/g_T$, each individual evaluation's $g_T$.  
The grey band gives the error assuming $\Delta g_T=0$, and the average $g_T$ (see Fig.\protect\ref{fig:gt}). The lattice evaluations from Refs. \cite{Green:2012ej,Bhattacharya:2013ehc} are indistinguishable.}
\label{fig:et}
\end{figure}
\end{center}
The impact on the extraction of $\epsilon_T$, of both the lattice QCD and experimental determinations of $g_T$ is regulated by the most recent limit \cite{Pattie:2013gka,Wauters:2013loa},
\begin{equation}
 \mid \epsilon_T  g_T \mid <  6.4 \times 10^{-4} ~~~~~~\mbox{(90\%CL)}.
\end{equation} 
Assuming no error on the extraction/evaluation of $g_T$,  yields $ \Delta \epsilon_{T, min}=6.4 \times 10^{-4}/g_T$. 
Since the errors on $g_T$ in both the lattice QCD and experimental extractions are affected by systematic/theoretical uncertainty, alternatives to the standard Hessian evaluation have been adopted in recent analyses \cite{Bhattacharya:2011qm} which are based on the R-fit method \cite{Hocker:2001xe,Gardner:2013aya}. 
By introducing the error on $g_T$,   
we obtain  $ \Delta \epsilon_{T} \geq \Delta \epsilon_{T, min}$.  The amount by which $ \Delta \epsilon_{T}$ deviates from the minimum error depends, however, on the relative error $\Delta g_T/g_T$ as well as on the central value of $g_T$, and on  $C_T$. 
We find that within the range of parameters extracted from our analysis of exclusive and semi-inclusive experiments,  knowing the tensor charge up to 
a moderate accuracy, $\Delta g_T/g_T \lesssim 20\%$, does not deteriorate the limits set by current experiments. This situation is illustrated in Fig.\ref{fig:et}, where we show $\epsilon_T$ vs. $\Delta g_T/g_T$, for the various determinations. 

In conclusion, the possibility of obtaining the scalar and tensor form factors and charges directly from experiment with sufficient precision, 
gives an entirely different leverage to neutron beta decay searches. 
While lattice QCD provides the only means to calculate quantities that are 
unattainable in experiment, for the tensor charge the situation is different. In this case, the hadronic matrix element is  the same which enters the DIS observables measured in precise semi-inclusive and deeply virtual exclusive scattering off polarized targets.
Most importantly the error on the elementary tensor coupling, $\epsilon_T$, depends on both the central value of $g_T$ as well as on the relative error, $\Delta g_T/ g_T$, 
therefore, independently from the theoretical accuracy  that can be achieved, 
experimental measurements are essential since they simultaneously provide a testing ground for lattice QCD calculations.



We are grateful to H. Avakian, A. Kim, S. Pisano, J. Zhang for details on the experimental extractions at Jefferson Lab,  and to L. Barr\'on Palos, M. Engelhardt, P.Q. Hung,  E. Peinado,  and D. Po\v{c}ani\'c for fruitful discussions. This work was funded by the Belgian Fund F.R.S.-FNRS via the contract of Charg?e de recherches (A.C.), by D.O.E. grant DE-FG02-01ER4120 (S.L.) and by NSF PHY-1205833 (S.B.). M.G.-A. is grateful to the LABEX Lyon Institute of Origins (ANR-10-LABX-0066) of the Universit\'e de Lyon for its financial support within the program "Investissements d'Avenir" (ANR-11-IDEX-0007) of the French government operated by the National Research Agency (ANR).

\bibliography{alu_bib}
\end{document}